\documentclass[pre,showpacs,floatfix,twocolumn]{revtex4}
\usepackage{graphicx}
\usepackage{amssymb}
\begin{document}

\title{Response to perturbations for granular flow in a hopper}
\author{John F. Wambaugh\footnote{Current address: National Center for Computational Toxicology, US EPA, Research Triangle
Park, NC 27711}}
\affiliation{Department
of Physics and Center for Nonlinear and Complex Systems, Duke University, Durham, NC 27708}
\email{wambaugh@phy.duke.edu}
\author{John V. Matthews}
\affiliation{Department of Mathematics, University of Tennessee at Chattanooga, Chattanooga, TN
37403}
\author{Pierre A. Gremaud}
\affiliation{Department of Mathematics and Center for Research in Scientific Computation, North
Carolina State University, Raleigh NC, 27695-8205}
\author{Robert P. Behringer}
\affiliation{Department of Physics and Center for Nonlinear and Complex Systems, Duke University, Durham, NC 27708}

\begin{abstract}

We experimentally investigate the response to perturbations of
circular symmetry for dense granular flow inside a three-dimensional
right-conical hopper. These experiments consist of particle tracking
velocimetry for the flow at the outer boundary of the hopper.  We are
able to test commonly used constitutive relations and observe new
granular flow phenomena that we can model numerically.  Unperturbed
conical hopper flow has been described as a radial velocity field with
no azimuthal component.  Guided by numerical models based upon
continuum descriptions, we find experimental evidence for secondary,
azimuthal circulation in response to perturbation of the symmetry with respect to gravity by tilting.  For
small perturbations we can discriminate between constitutive
relations, based upon the agreement between the numerical predictions
they produce and our experimental results.  We find that the secondary
circulation can be suppressed as wall friction is varied, also in
agreement with numerical predictions.  For large tilt angles we
observe the abrupt onset of circulation for parameters where
circulation was previously suppressed.  Finally, we observe that for
large tilt angles the fluctuations in velocity grow, independently of
the onset of circulation.

\end{abstract}



\date{\today}

\maketitle

\section{Introduction}

Many practical processes involve the flow of dense granular materials.
Applications range from food grains, ores, and coal to pharmaceutical
powders.  To predict flows of dense granular states, one would like a
homogenized/continuum model that reflects the complexity of
grain-scale interactions but can be applied at much larger scales.
Continuum approaches frequently describe granular materials using
elasto-plastic models in which the material can support shear stresses
up to a certain value before yielding and deforming irreversibly.  In
the simplest case, this approach amounts to Coulomb's law of friction.

The most common yield criterion is the Mohr-Coulomb criterion, which
proposes a linear relationship between the magnitudes of normal stress
$\sigma$ and shear stress $\tau$ in two-dimensions.  Because the
choice of axes dictates the relative magnitude of the normal and shear
stresses, the range of possible axes maps out a circle in the space of
normal and shear stress --- the ``Mohr circle".  The diameter of this
circle is proportional to the magnitude of the stresses
\cite{Nedderman92}.

In the normal-shear stress space, the Mohr-Coulomb yield criterion is
the straight line $\tau = \mu \sigma + c$ where $\mu$ is the
coefficient of friction and $c$ is the cohesion.  It follows for
cohesionless granular materials, the case we consider here, that the
yield criterion must pass through the origin.  The stress within the
material, and hence the diameter of the Mohr circle, can increase only
until the yield criterion is tangent to the circle.  The point of
contact then indicates an axis of a coordinate frame in which the
ratio of shear to normal stress (equivalent to the mobilization of
friction) is maximized.  In this two-dimensional analysis of a
three-dimensional material, the direction of the maximally mobilized
shear stress and the third, neglected, axis span the ``mobilized
plane" along which motion of the granular material is most likely in
response to stress \cite{Nedderman92}.

There are several approaches that have taken for generalizing the Mohr
circle picture to three dimensions.  An example is the Matsuoka-Nakai
criterion.  By geometrically averaging the three planes obtained by
considering pairs of dimensions a ``spatially mobilized plane" is
obtained.  This is similar to the generalization of the Tresca failure
criterion for metals to three dimensions by von Mises
\cite{MatsuokaNakai85}.  Tri-axial experiments support the validity of
the Matsuoka-Nakai criterion, indicating that it may capture the
average behavior of a granular material \cite{Matsuoka89}.

Unfortunately, the mathematics of continuum descriptions of granular
materials are rife with instabilities, ill-posedness, and complex
non-linearities \cite{Schaeffer90a}. This means that evaluating the
response of granular materials to perturbations that cause unusual
behavior in continuum models is necessary to determine the physical
relevance of such approaches.

The right-conical hopper provides an ideal test system for granular
flow in that it is not only an experimentally-accessible system, but
also because the behavior is commonly assumed to tend to a
steady-state with known velocity and stress field solutions derived
from soil mechanics.  The so-called Jenike radial solutions describe
the velocity and stress fields in an infinite hopper as self-similar
functions of radial position alone, without non-radial velocity
components \cite{Jenike61}.  In two dimensions, such soil mechanics
descriptions of wedge hoppers have been studied extensively both
experimentally \cite{Bazant04a,Horluck01,Trevino98a,Chou02} and
theoretically \cite{Nedderman85b,Nedderman96}.  Three-dimensional,
right-conical hopper flow has been studied through experiments and
discrete element simulations in the past\cite{Baxter93,Ristow98}, but
these studies did not make direct comparison to soil mechanics
predictions.

Recent numerical work has taken advantage of the self-similar nature
of the solution even in relatively general three-dimensional
geometries such as ``pyramidal" hoppers.  By perturbing the geometry
of the simulated hopper, the response of the continuum description can
be examined.  In particular, numerical work has found that secondary,
non-radial circulation currents arise.  These currents depend sharply
upon model parameters in non-trivial ways
\cite{Gremaud03,Gremaud04,Gremaud06}.

We make quantitative experimental observations of a right-conical
granular hopper to determine if these numerically-predicted behaviors,
the result of non-linearities in the continuum description of granular
materials, have physical significance.  We investigate both the
influence of tilt-induced perturbations and changes in boundary
roughness.  By comparing with numerical predictions we investigate the
ability of different constitutive relations to correctly model
experimentally observed phenomena.  Since we obtain localized
time-resolved velocities, we also identify useful statistical
characterizations of flowing dense granular matter.

\section{Soil Mechanics Approaches to Granular Matter}

Before turning to the experiments, we give a brief summary of the
relevant soil mechanics constitutive relations.  At issue are the
determining equations for the stress tensor.  Force balance gives six
equations for the nine components of the three dimensional stress
tensor.  To close this system of equations we need additional
constitutive relations describing the yielding, plasticity and flow of
the material \cite{Gremaud04}.  Consideration of the velocity field
$v$ describing the flow introduces three more unknowns bringing the
necessary number of additional constraint equations to six.

We use Levy's flow rule to introduce five constraints
\cite{Jenike87,Nedderman92}.  Missing is a final constitutive relation
describing the stress sufficient to cause plastic rearrangement, or
yield, within the material \cite{Howell99}.

In metal plasticity, the von Mises failure criterion is commonly used
as a constitutive relation.  The von Mises yield surface in stress
space, however, is independent of the total pressure:
\begin{equation}
\left(\sigma_1 - \sigma_2\right)^2 + \left(\sigma_2 - \sigma_3\right)^2 +\left(\sigma_3 - \sigma_1\right)^2 = c^2
\end{equation}
where $\sigma_1$, $\sigma_2$, $\sigma_3$ are the principal stresses
denoted in order of magnitude and $c$ is a constant.  This is
necessarily at odds with the Coulomb nature of granular failure.
Replacing the constant term on the right-hand side by a term involving
the principal stresses generalizes the von Mises criterion to granular
materials by the introduction of a pressure-dependent yield surface:
\begin{equation}
\left(\sigma_1 - \sigma_2\right)^2 + \left(\sigma_2 - \sigma_3\right)^2 +\left(\sigma_3 - \sigma_1\right)^2 = 6\sin{\theta_s}^2I_1^2
\end{equation}
or equivalently,
\begin{equation}
9I_1^2 + 2I_2 = 6\sin{\theta_s}^2*I_1^2
\end{equation}
where the three stress invariants are $I_1 =
\frac{1}{3}\left(\sigma_1+\sigma_2+\sigma_3)\right)$ (equivalent to
the isotropic pressure), $I_2 = -
\left(\sigma_2\sigma_3+\sigma_3\sigma_1+\sigma_1\sigma_2)\right)$ and
$I_3 = \sigma_1\sigma_2\sigma_3$ and $\theta_s$ is the internal angle
of friction of the sand \cite{Nedderman92,MatsuokaNakai85}.  This
``granular von Mises" condition is commonly used as the missing
granular constitutive relation.

As an alternative, Matsuoka and Nakai \cite{MatsuokaNakai85} apply the
Mohr-Coulomb failure criterion $|\sigma_{jk}| \leq
\tan\phi_i\sigma_{kk}$ for a cohesionless material to obtain, for each
pair of axes $j$ and $k$, a Mohr circle with radii $R_i =
\left(\sigma_j - \sigma_k\right)\sin\theta_s$ and a yield criterion
that makes an angle $\phi_i$ at the origin such that $\cos{\phi_i} =
|\sigma_j-\sigma_k|/R_i$.  They argue for the following constitutive
relation:
\begin{equation}
\tan{\phi_1}^2 +\tan{\phi_2}^2 +\tan{\phi_3}^2 =c
\end{equation}
which they show to be equivalent to:
\begin{equation}
I_1*I_2/I_3 = c
\end{equation}

Gremaud et al. have computed flows in relatively general geometries
for both of the above plasticity models \cite{Gremaud04,Gremaud06}.
The authors model granular flow in a vertically oriented infinite
right-conical hopper described in spherical coordinates where $r$ is
the distance from the origin, which is at the tip of the cone,
$\theta$ is the angle measured away from the axis of symmetry and
$\phi$ is the azimuthal position about the symmetry axis.  The authors
vary the wall angle azimuthally about the constant angle $\theta_w$ as
$\theta = \theta_w + \epsilon \cos{m\phi}$.  The case $m=1$
corresponds to elongating the cross-section of the hopper in a manner
that is roughly similar to the effect of tilting the hopper with
respect to gravity.  Assuming $\epsilon$ is small (between $-5^\circ$ and
$5^\circ$ for a tilted hopper), the velocity can be expanded as $v = v^0 +
\epsilon v^1 + ...$ giving the radial and azimuthal components of the
velocity as:
\begin{eqnarray}
v_r &\approx& v^0_r + \epsilon v_r^1\nonumber\\
v_\phi &\approx& \epsilon v_\phi^1
\end{eqnarray}
where an unperturbed ($\epsilon = 0$) hopper has Jenike radial flow.

\begin{figure}
\centering
\includegraphics[width=3.375in,clip=]{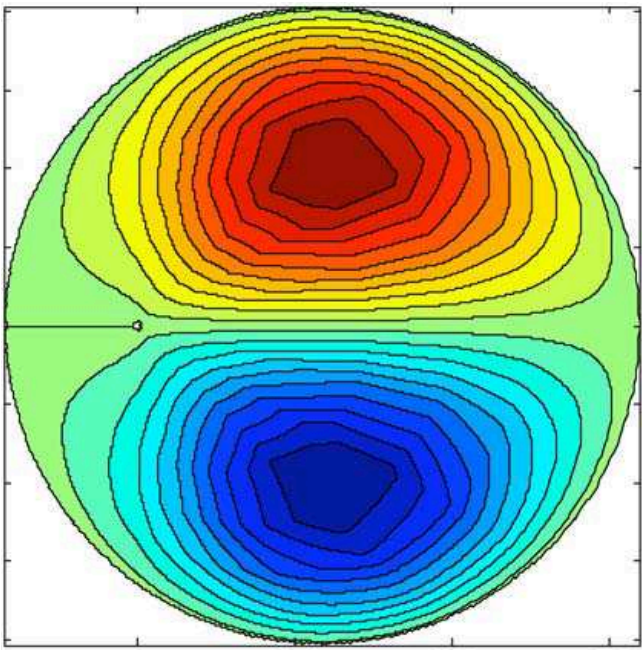}
\caption{(Color online) Numerical simulations of the velocity stream functions for tilted hoppers predict the rise of two secondary circulation cells, rotating in opposite directions.  Darker tones represent greater velocity.  In the top half of the figure the circulation is clockwise while in the lower half the circulation is counterclockwise.  The strength and shape of the cells depend on the angle   of tilt, angle of internal friction, and coefficient of wall   friction.  The azimuthal angle $\phi$ is taken to be zero on the line of symmetry between the two cells.}
\label{numericalfield}
\end{figure}

Two main observations were drawn from the numerical experiments of
Gremaud et al.  First, when vertical axi-symmetry is broken, secondary
circulation takes place, i.e., $v_\phi$ becomes non-zero and leads to
two counter-rotating cells perpendicular to the tilt axis, as in
Fig.~\ref{numericalfield}.  Second, the computed circulation currents
were much larger for a Matsuoka-Nakai material than for a von Mises
one.

The numerically predicted components of velocity $v^h_\phi$ and
$v^h_r$ depend upon the choice of model parameters and constitutive
relation.  When using the Matsuoka-Nakai relation, $v^h_\phi$ varies
strongly as a function of wall friction, $\mu_w$, ranging from the
same order as the radial flow to three or more orders of magnitude
smaller.  Similar behavior is seen when the von Mises granular
criterion is used, but the effect is smaller by orders of magnitude
\cite{Gremaud06}.

The numerical model assumes mass flow, where all the grains are moving
and there are no stagnant regions within the hopper.  For this reason,
simulations were not conducted above $\mu_w = \tan(\theta_s) = 0.5$,
where the angle of friction of the simulated material was $\theta_s =
30^\circ$ \cite{Gremaud03}.

\section{Methodology}

\begin{figure}
\centering
\includegraphics[width=1.8in]{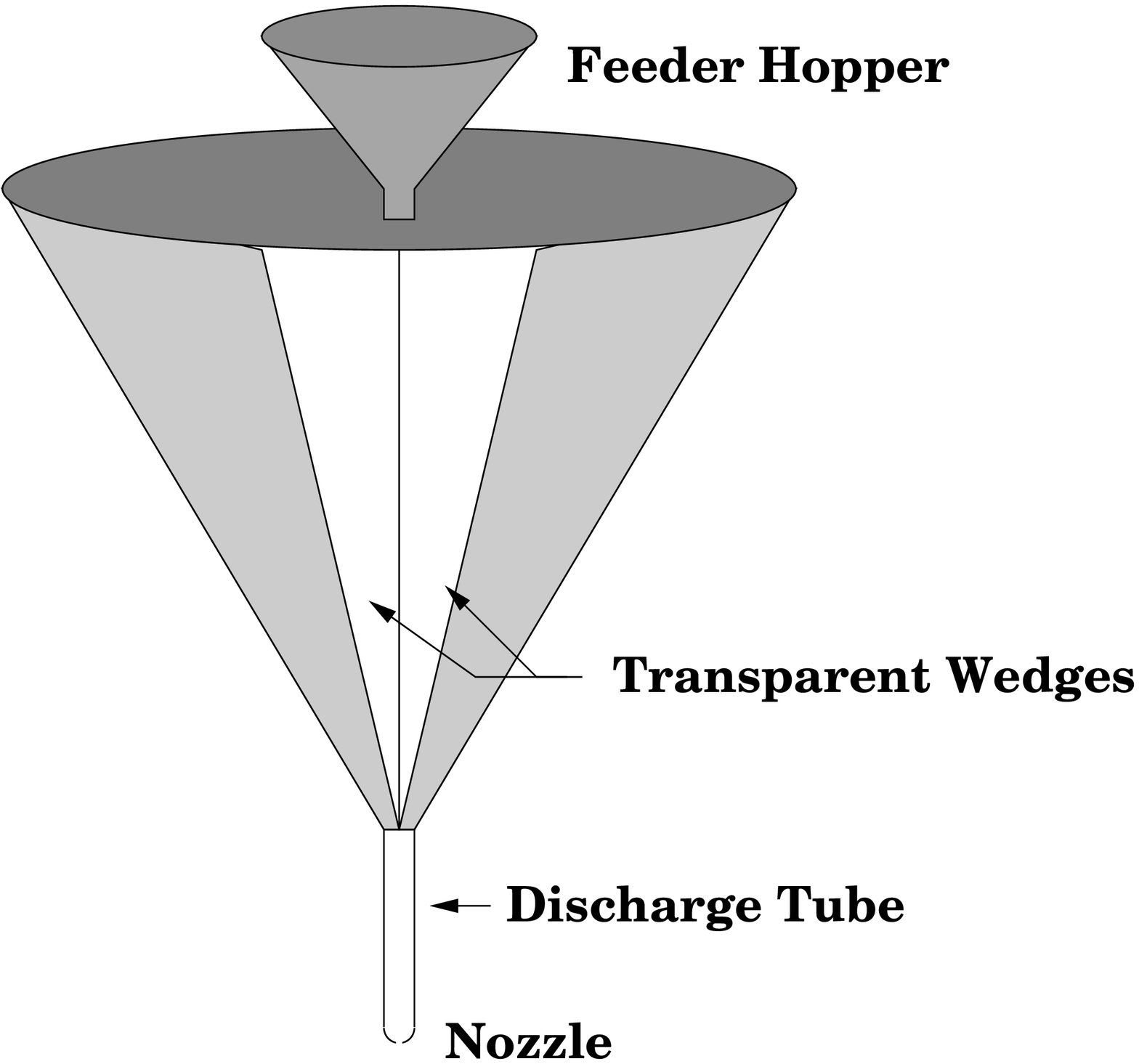}
\includegraphics[width=1.2in,clip=]{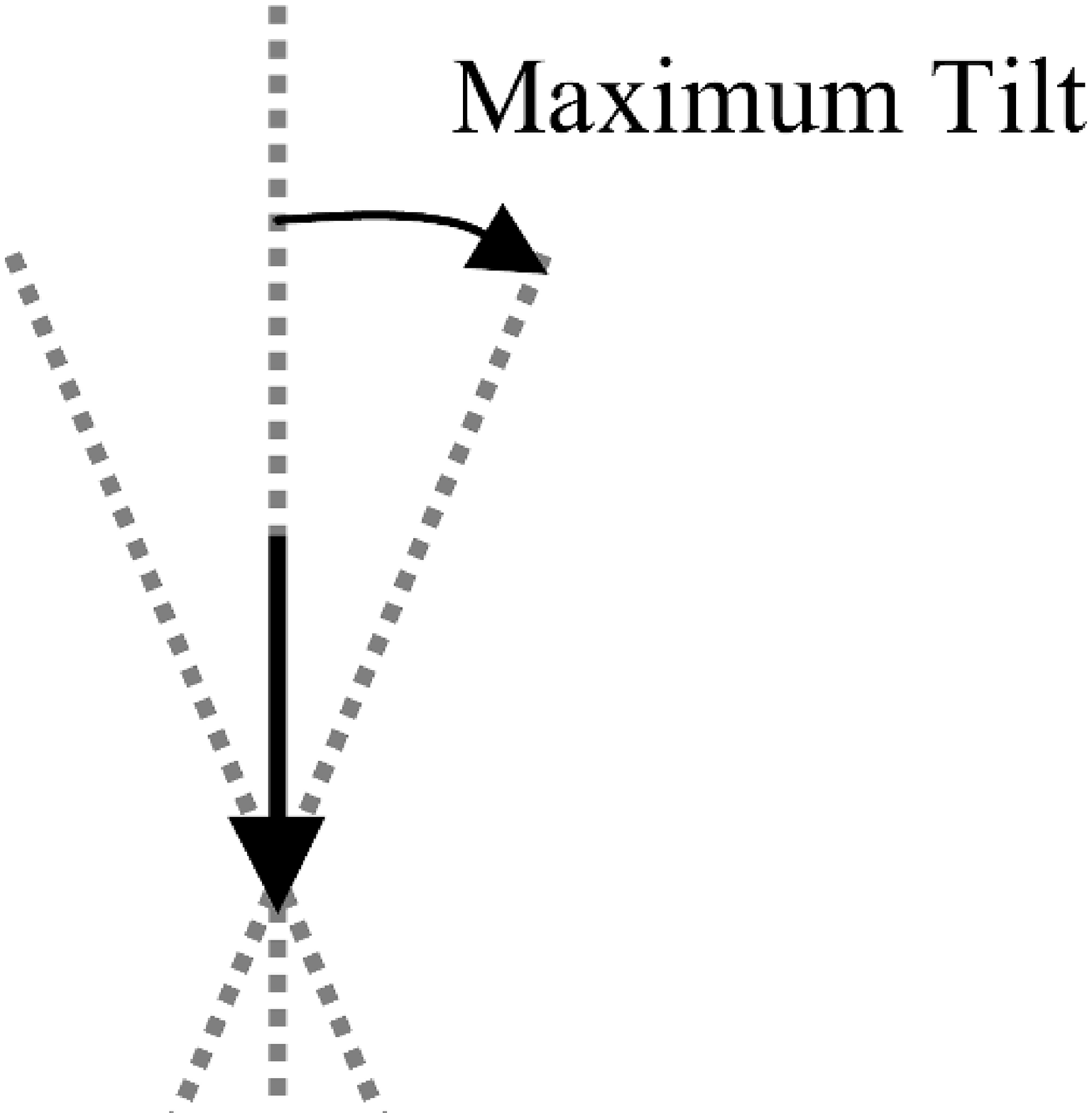}
\caption{Schematic of hopper in which we observe flowing sand.  To perturb the flow we tilt the hopper up to 23 degrees.}
\label{hopper}
\end{figure}

Our experimental test hopper is an approximately-conical, regular
polyhedron composed of fourteen brass wedges ($\mu_w = 0.34$ when
bare) and two clear Plexiglas wedges ($\mu = 0.51$).  The hopper, depicted in Fig.
\ref{hopper}, is $\sim45$ cm wide at the top and the walls are angled
at $\theta_w = 24^\circ$.  Flow is slowed by a nozzle with a $0.6$ cm diameter opening to $\sim6$ g/s and
the nozzle is screened from the hopper by a $3$ cm diameter tube that is sufficiently long ($\sim20$ cm) for the influence of the nozzle to be neglected.  Sand runs from the opening at the bottom of the hopper into the discharge tube, with foam wrapped around the joint to prevent spillage.
It takes $\sim20$ minutes for the hopper to drain if not refilled.  A
second, feeder hopper is placed on axis at the top to provide
refilling from an approximate point source $\sim14$ cm within the hopper.  The feeder hopper is refilled by hand as needed and keeps the main hopper nearly-fixed at roughly one third full.  By
feeding from the small hopper we maintain a well-controlled
preparation of the sand in the primary hopper.

We note that the 16-sided cross section of the hopper involves some perturbation from
circular symmetry.  However, we estimate that these are 15 to 20 times smaller than the
typical perturbation that we achieve by tilting the hopper.  In addition, we expect that any small flows induced by these perturbations would not have a substantial swirling
component.

In these experiments we use Ottawa sand (ASTM C-190).  The sand
(coefficient of friction $\mu_{sand} = 0.42$, density $1.64 \pm 0.01$ g/mL, and diameter $420 \leq d
\leq 595$ $\mu m$) is mixed with a small quantity of tracer sand that
has been dyed dark red with instrument ink and repeatedly baked dry.

The entire hopper can be physically tilted, perturbing the geometry with respect to gravity.  The feeder hopper tilts with the system, staying on the axis of the main hopper.  A digital (CCD) camera, bolted to the side of the hopper, tilts with the hopper and can be
aligned with the Plexiglas ``windows" in one of two configurations so
that images showing the passage of sand may be recorded with a
computer.  The hopper is tilted so that the windows are on the axis of
the tilt, where $v_\phi$ is predicted to be maximal.  A plumb line is
hung within the hopper and imaged to determine the direction of
gravity in the frame of the camera for each tilt.

Since the hopper is made of flat wedges shaped into a cone, each wedge has a different orientation relative to the camera.  To account for the tilt of the camera
relative to the orientation of the window wedges we mount the camera
two different ways.  In the first, we align the camera so that the
tilts of the two windows are equal but opposite in orientation
relative to the camera.  In this arrangement, any geometric effect
caused by the tilt of one window should be reversed in the other
window.

We use the second camera arrangement when we line the walls of the
hopper with materials with different coefficients of friction. To
minimize the impact of the viewing window when lining the hopper we
cut a hole no bigger than the size of the field of view of the camera
into the liner.  In this case we align the camera flush with just one
window of the hopper.

The coefficient of friction, $\mu_w$, of a particular liner material
is determined by using a weight and pulley system to drag a tray covered
with the material across a
bed of Ottawa sand.  We capture the motion of the tray using a
high-speed digital camera and analyze the images to determine the
acceleration of the tray.  For each run there is a period where the
forces on the sled are roughly constant and we perform a linear fit to
the velocity in this region to determine acceleration.  By varying the
load applied to the tray as well as the weight acting through the pulley we can account for systematic frictional forces and
deduce the sliding coefficient of friction for sand on the liner
material.

Data sets for determining the velocity field typically consist of 1000
frames taken once every two seconds --- covering roughly thirty-three
minutes of flow.  For a given combination of tilt and wall material,
from two to six data sets are used to determine each plotted point for
$v$.

\begin{figure*}
\centering
\includegraphics[width=3.375in,clip=]{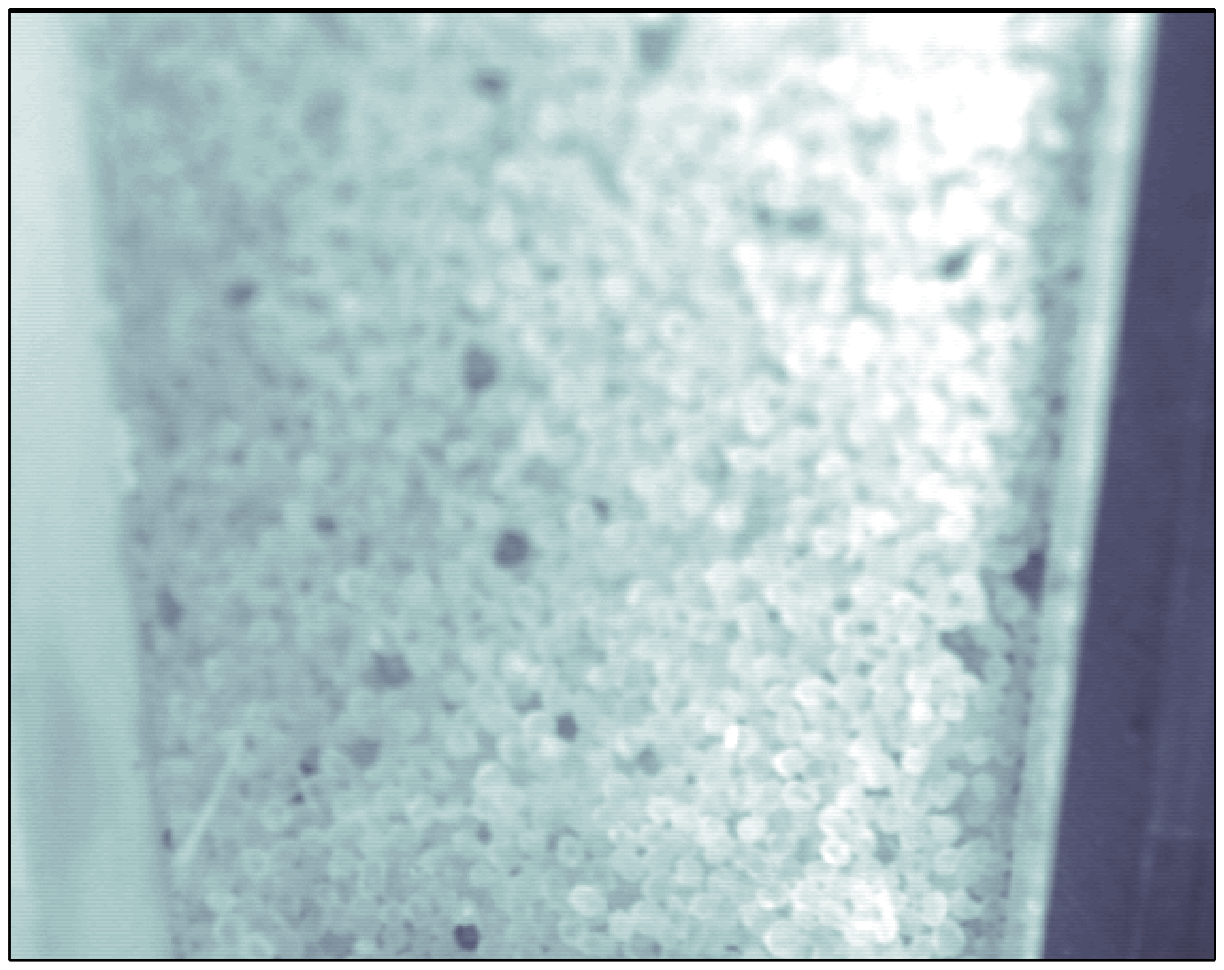}
\includegraphics[width=3.375in,clip=]{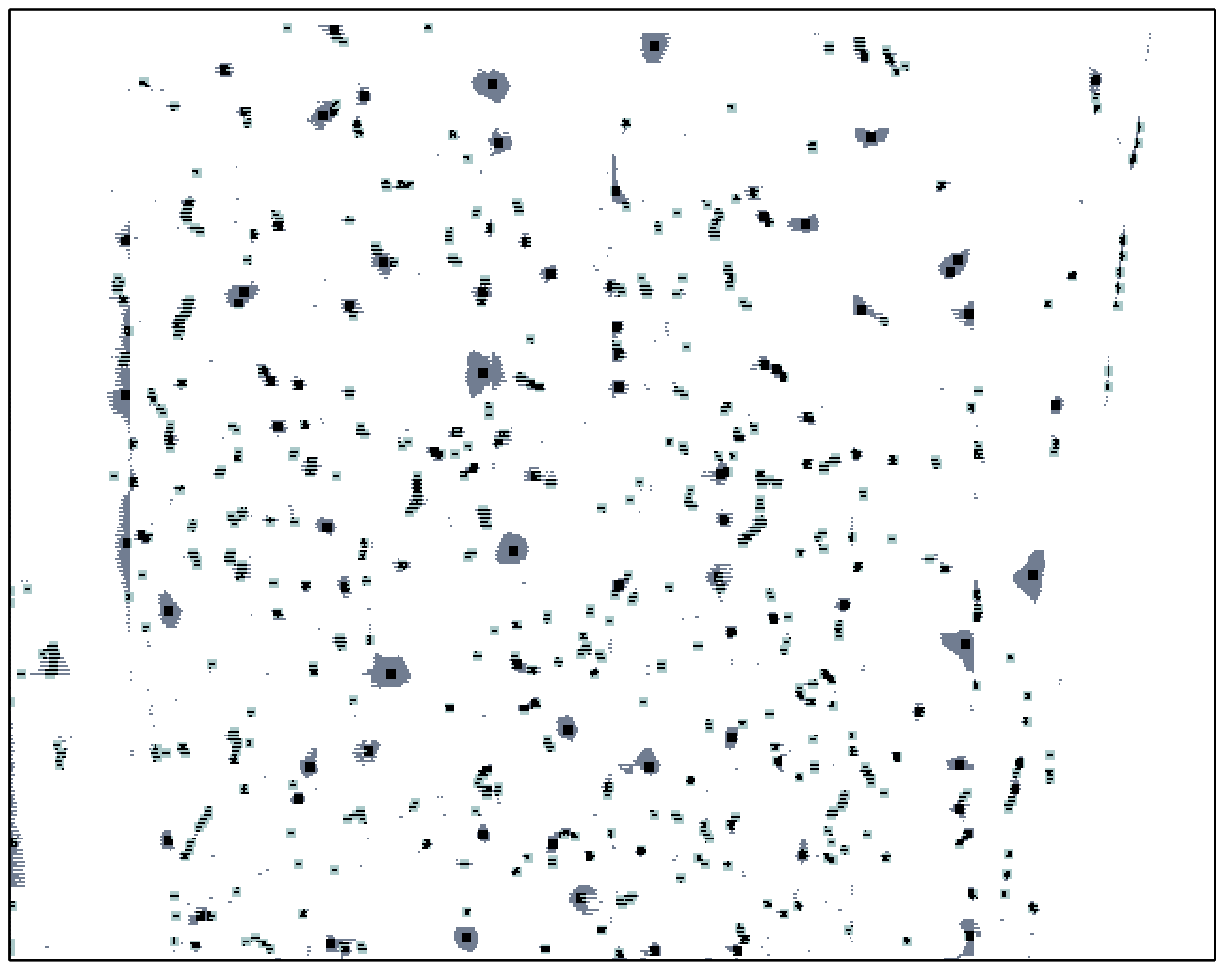}
\includegraphics[width=3.375in,clip=]{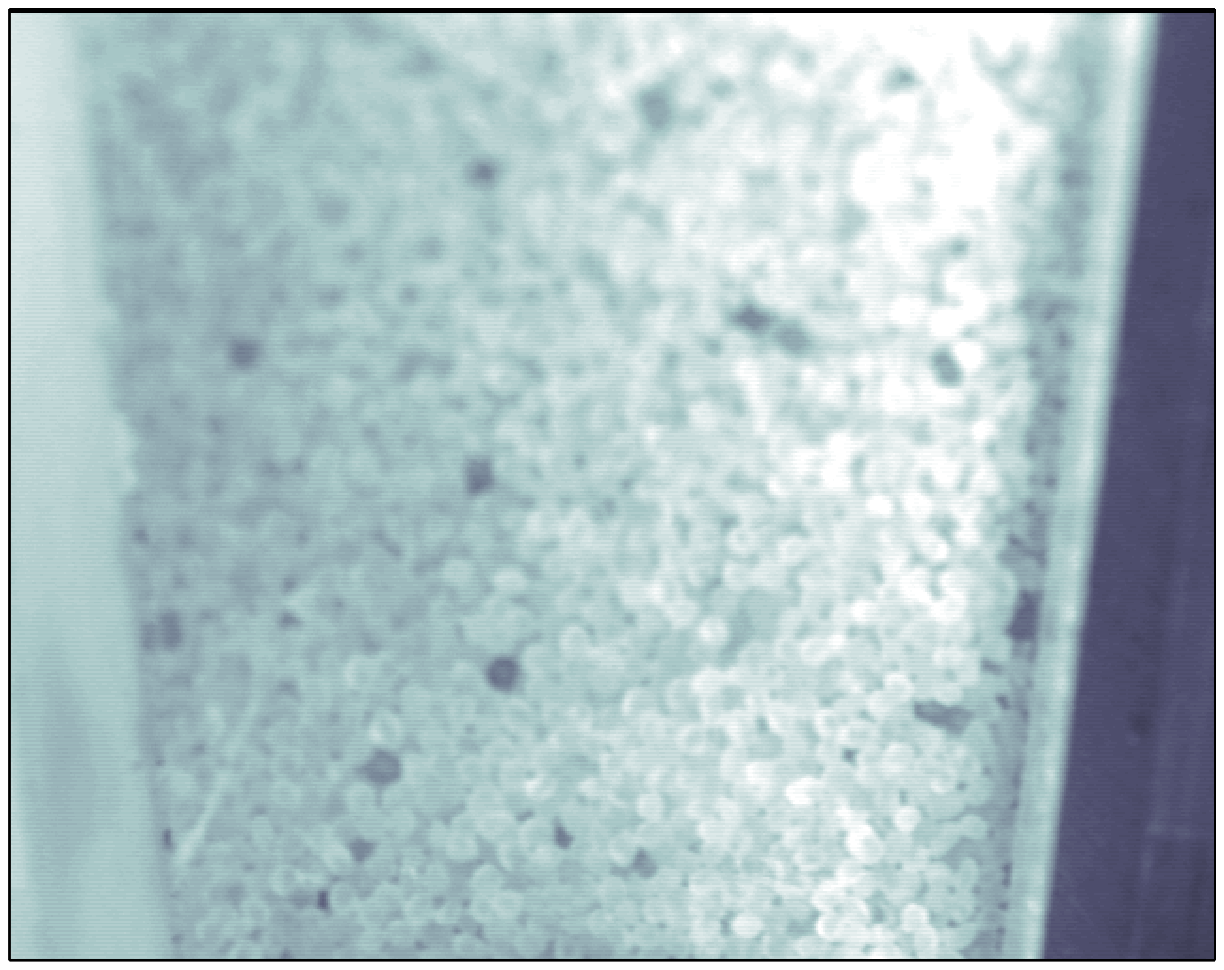}
\includegraphics[width=3.375in,clip=]{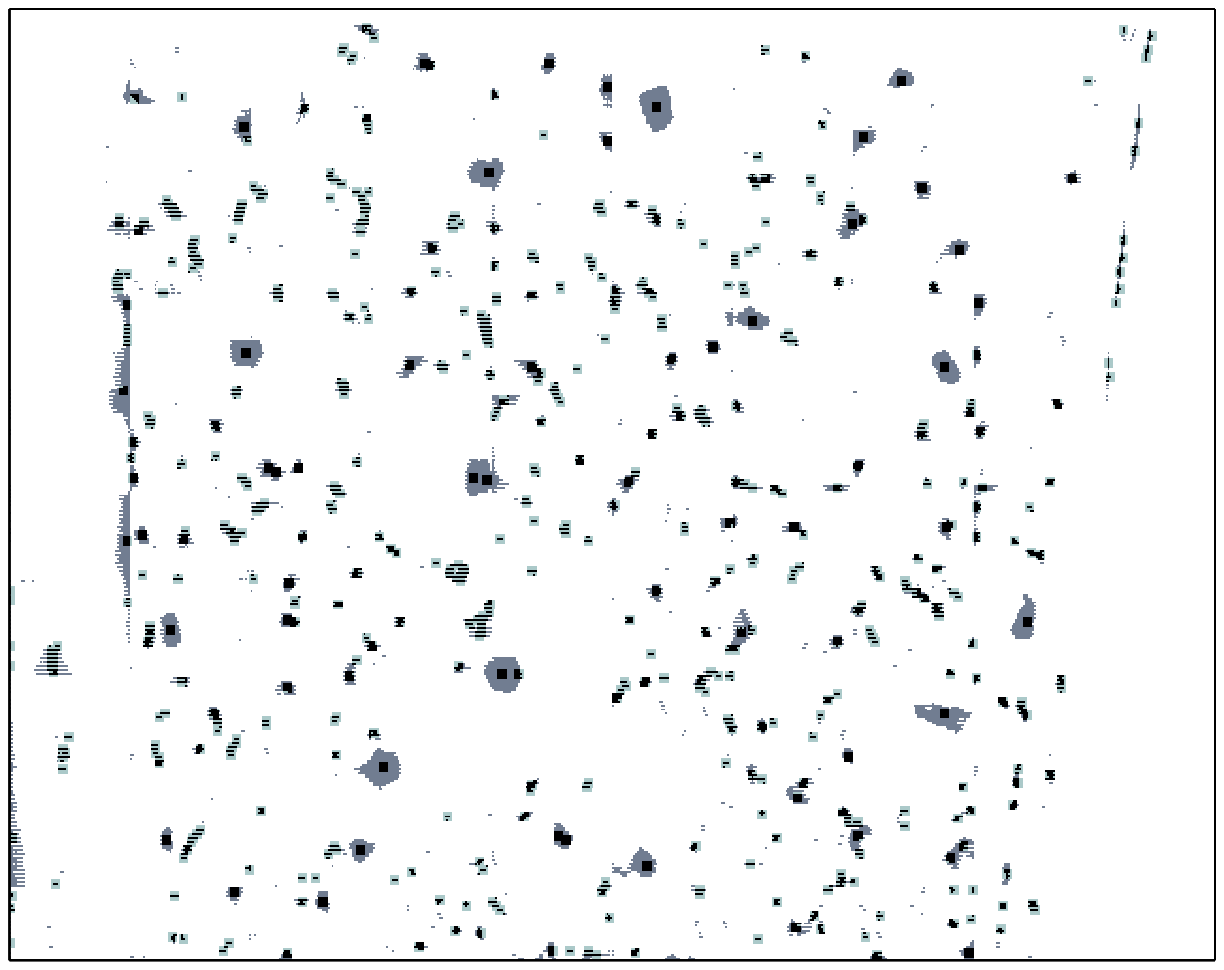}
\caption{Velocities are determined from identifying tracer particles
and regressing over twenty-one consecutive frames to assign a velocity
to the eleventh frame.  Shown here are the actual frames ({\bf left})
and thresholded tracer bitmaps ({\bf right}) for a given frame ({\bf
top}) and the twentieth frame later ({\bf bottom}), corresponding to
$40$ s of experimental observation.  The dark dots indicate centers
assigned to large tracer particles.}
\label{valgo}
\end{figure*}

As illustrated by Fig.~\ref{valgo}, particle tracking is performed by
thresholding each frame into a binary intensity image where `ones'
identify pixels with intensity that is a tunable number of standard
deviations away from the local mean intensity in the region of the
pixel.  The thresholding is done by region to account for any
non-uniform lighting.  Clusters of `ones' larger than a tunable size
are then identified as the dark, tracer particles amidst the lighter
sand, although occasionally bright or reflective sand grains are also
resolved and tracked.  Typically a dozen or more grains are tracked in
a frame.

\begin{figure} \centering
\includegraphics[width=3.375in]{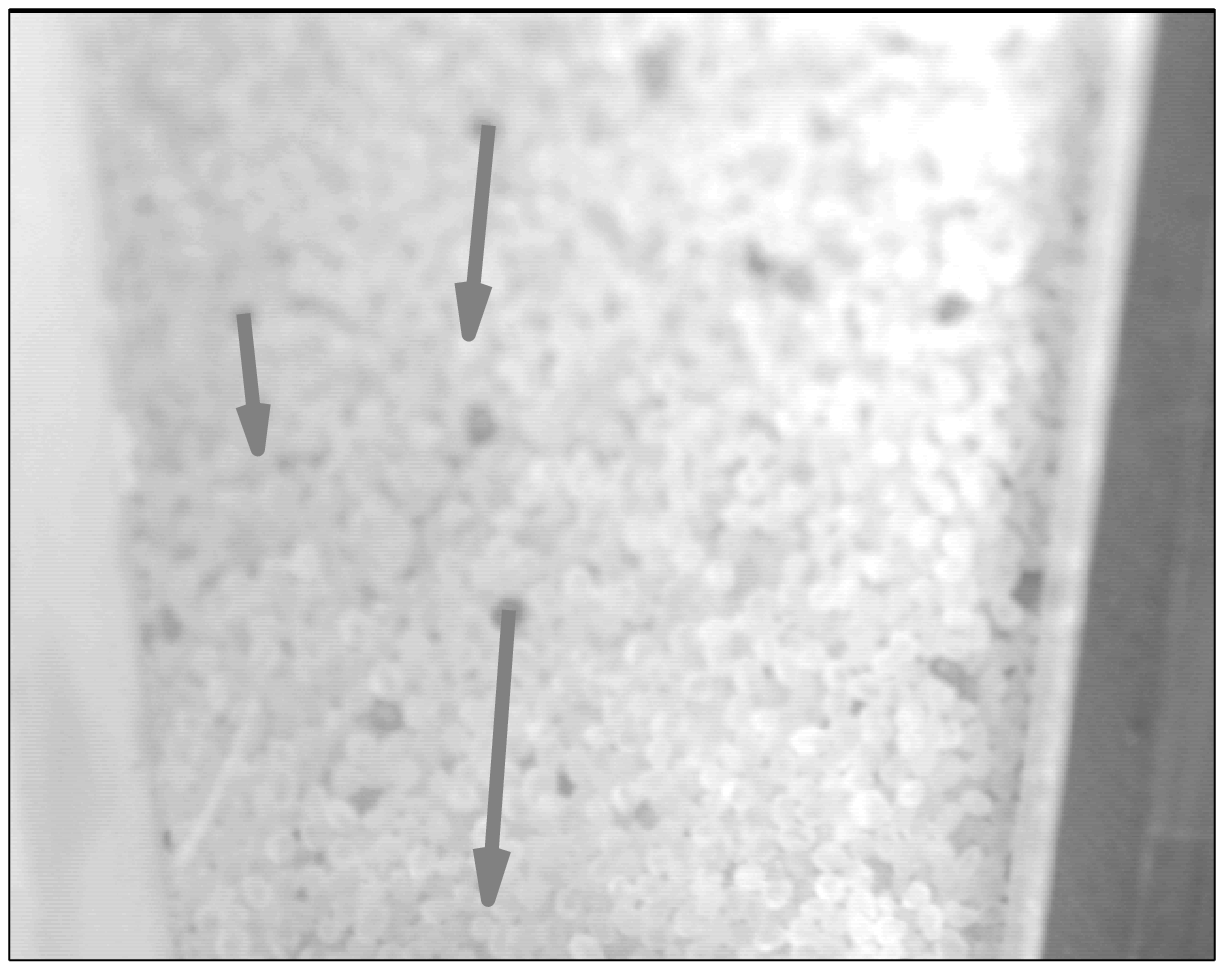}
\caption{The velocities assigned to the eleventh frame in the sequence
depicted in Fig.~\ref{valgo} indicate that in that frame only three
tracer particles had sufficiently-long unambiguous tracks to be used.}
\label{assignedv}
\end{figure}

Since the grains are slightly irregular, the cluster identifying a
given tracer grain changes as the particle rotates along the window
surface.  Typical tracer particles are roughly $20\times20$ pixels in
size, though smaller dark or bright spots are tracked and used if
possible.  Each cluster is assigned a ``center of mass" determined
from the spatial distribution of its pixels, and then is identified
with the nearest center of mass in the previous frame to construct a
track of particle position in consecutive frames.  If the nearest
center of mass is outside a tunable maximum radius, the identification
is rejected and a new particle track begins.  Once a track is
twenty-one frames long, a velocity is calculated using a least-squares
fit and the interpolated velocity is assigned to the location of the
tracer in the eleventh frame, as in Fig.~\ref{assignedv}, where out of
the dozens of possible tracers only three are being used.  If the
velocity is zero, indicating an edge of the window is being tracked,
it is ignored.  Particle tracks begin and end as tracer particles move
into and out of the field of view of the camera due to the overall
flow and motion towards and away from the window --- the mean track
length is roughly 26 frames.  In this manner we generate a
time-integrated velocity field by binning velocities into regions by
tracer location over the length of the run.

\begin{figure}
\centering
\includegraphics[width=3.375in,angle=270]{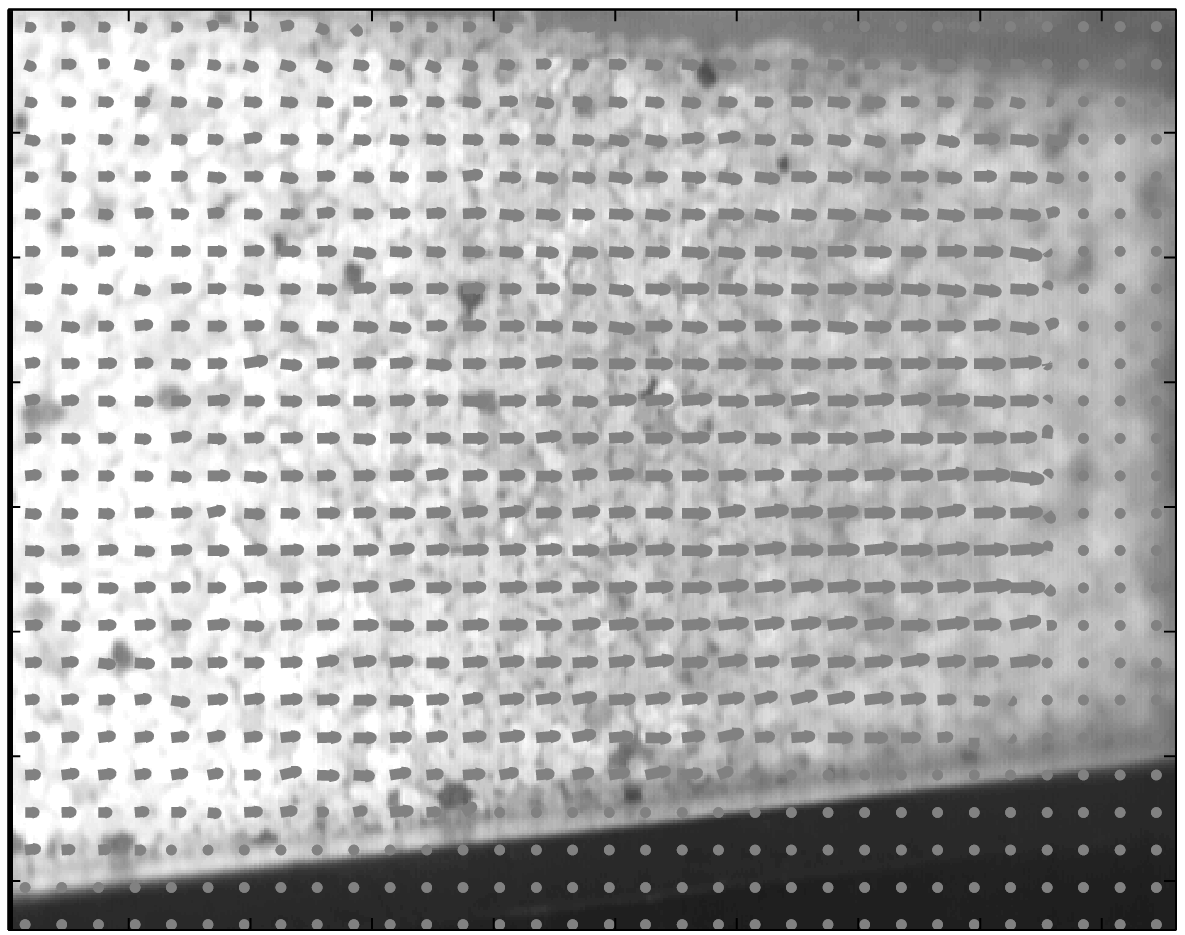}
\caption{A time-averaged velocity field is generated by spatially
binning the velocities for every frame.}
\vspace{0.1in}
\label{field}
\end{figure}

Once a velocity field is generated for a particular combination of
tilt angle and wall friction, as in Fig.~\ref{field}, we can calculate
the ratio $v_\phi/v_r$.  If we have recorded using two, off-setting
windows, we calculate the ratio separately for the two windows.  We
first linearly fit both components of velocity horizontally (fixed
$r$) across the velocity field to the distance from the middle of the
image to compensate for the slight tilt of the windows away from the
camera.  We interpolate using this fit to find the velocity components
nearest the middle of the field of view of the camera from both
windows, corresponding to $\phi = \pi/2$ ($\phi = 0$ is in the plane
of the tilt).  If we are using a single window, we perform a quadratic
fit across the velocity field to account for $\phi$ variation (a
higher order effect in the two-window case).  We again interpolate the
velocity components at $\phi = \pi/2$.

We find both the azimuthal and radial components of the velocity to
vary radially as $1/r^2$, in agreement with mass conservation and the
Jenike solutions.  Given that both components vary as $1/r^2$, the
ratio of the two velocity components, $v_\phi/v_r$, should be
independent of $r$.  Linear fits to $1/r^2$ of the two velocity
components are made and for each radial bin of the vector field the
ratio of the $v_\phi$-fit to the $v_r$-fit is calculated and the mean
is recorded as the observed ratio.  Error bars are established using
the standard deviation of these ratios.

\section{Results}

We find evidence for a non-radial component of the velocity field as a
function of hopper tilt and wall friction.  We also find that small
perturbations of geometry, such as slight misalignment of the wedges
forming our test hopper, causes a small amount of apparent
secondary-circulation.  With careful alignment and balancing of our
untilted experimental hopper we were able to reduce, but not
completely eliminate, this apparent non-radial flow even for the
untilted case.  In order to compensate for the observed small
non-radial component of the flow, we rotate the experimental images to
a frame in which the untilted hopper flow is perfectly radial.

\begin{figure}
\centering \includegraphics[width=3.375in]{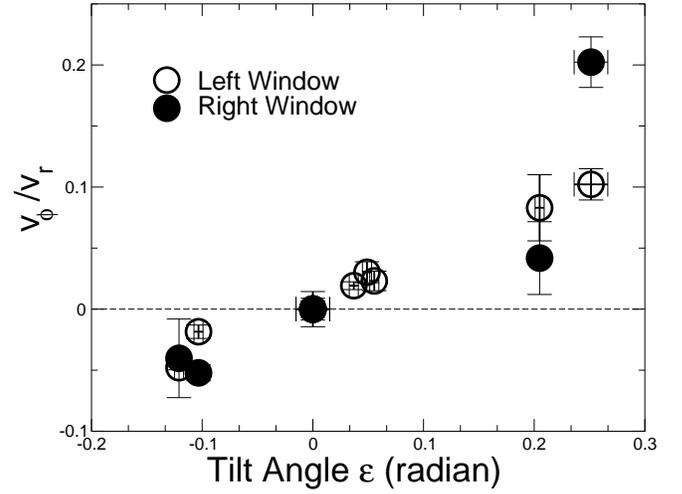}
\caption{The ratio of azimuthal to radial velocity for the right- ($\bullet$) and left-hand
($\circ$) windows for a bare ($\mu_w =
0.34$) hopper.}
\label{circulation}
\end{figure}

As the measurements from our test hopper shown in
Fig.~\ref{circulation} indicate, in a bare brass hopper ($\mu = 0.34$)
secondary circulation does in fact follow linearly with tilt angle for
small perturbations ($\epsilon < 0.15$).  At larger angles, the
dependence appears to depart from linear, which might be expected,
since higher-order terms should eventually become important.  Note
that the ratio of azimuthal to radial velocity, \(v^h_\phi/v^h_r\),
follows systematically for tilts in both directions, indicating that
our observations are not an artifact.

\begin{figure}
\centering
\includegraphics[width=3.375in,clip=]{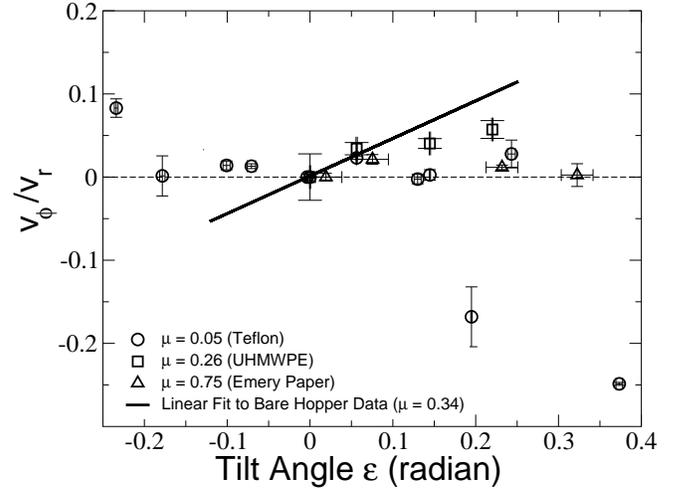}
\caption[Comparison of bare hopper to results for various wall
frictions]{Comparison of the ratio of azimuthal to radial velocity for
Teflon ($\mu_w = 0.05$ --- $\bigcirc$), UHMWPE ($\mu_w=0.26$ ---
$\Box$) and Emery paper ($\mu_w = 0.75$ --- $\triangle$) to a linear
regression of the bare ($\mu_w = 0.34$) hopper observations ({\bf
solid line}).}
\label{frictions}
\end{figure}

Numerically it is possible to predict $v_\phi^h / v_r^h$ as a function
of $\mu_{sand}$, $\theta_{wall}$ and $\mu_{wall}$, although there are
many domains where the predicted magnitude of secondary circulation is
too small to be observed.  For our system, wall friction is the most
readily tunable parameter and we observed the flow for three
additional wall frictions, as indicated by Fig.~\ref{frictions}.

By choosing the appropriate material, we were able to examine circulation
in regimes where azimuthal flow would be suppressed.  One such domain
is that of low wall friction values, (such as Teflon --- $\mu =
0.05$).  In the linear region ($\epsilon < 0.1$) there is no
detectable secondary circulation, within experimental resolution, when
the hopper is lined with teflon.  As the tilt increases however, a strong azimuthal component is observed in the opposite direction of the flow in the bare hopper.
The tilt angle for the onset of flow showed some variation, possibly
indicating hysteretic effects or some other
influence beyond the controls of the experiment.

As wall-friction is increased beyond $\mu_{critical}$, soil mechanics
models of granular flow predict a transition from mass-flow --- where
all the material in the hopper is moving --- to funnel flow where
there are non-moving, stagnant regions at the wall.  However, when we
line the hopper with Emery paper, which has a higher coefficient of
friction $\mu = 0.75$ than the interparticle friction {\bf $\mu_{crit}
= \tan{\theta_{sand}}$} we
continue to observe some radial flow at the window, although there is
no azimuthal component within measurement error.

Although extreme values of wall friction induce large changes in flow,
we find that the flow is relatively robust to small variations in wall
friction.  Changing the wall friction slightly from the bare hopper
(from $\mu_w = 0.34$ to $0.26$), by lining with low-friction
Ultra-High Molecular Weight Polyethylene (UHMWPE), does not
qualitatively change the secondary circulation.

\section{Comparison with Soil Mechanics Predictions}

\begin{figure}
\centering
\includegraphics[width=3.375in]{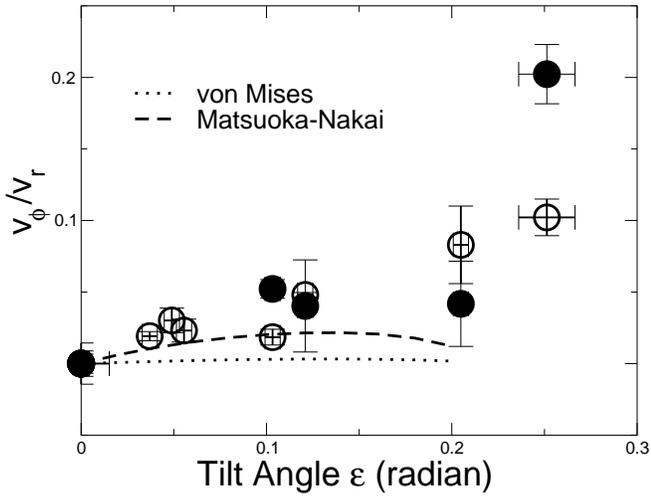}
\caption{Ratio of azimuthal to radial velocity for a bare hopper as a function
of hopper tilt angle $\epsilon$ as imaged through the right
($\bullet$) and left ($\circ$) windows of the test hopper, rotated to
the frame where the untilted flow is entirely radial and mapped so that all tilt angles are positive.  Lines indicate
numerical predictions for von Mises (dotted) and Matsuoka-Nakai
(dashed) plasticity models.}
\label{comparisonbare}
\end{figure}

The dashed and dotted lines in Fig.~\ref{comparisonbare} show
azimuthal to radial velocities ratios as calculated for varying tilts
in a bare hopper using a spectral method \cite{Gremaud06}.  We examine the ratio at the point in the hopper where we observed experimentally.  As the tilt increases, we predict the onset of stronger and stronger circulation.  In
this respect the predicted behavior the simulation and experiment are similar for small tilt angles.

As the tilt increases further, we see in the simulations that the circulation cells move relative to the point on the hopper wall where we have observed the flow experimentally.  This movement of the circulation cells puts the observation area at a point where the rotational flow is much smaller and approaches zero.  This decrease in the flow is not observed experimentally.

We find that the values predicted using the Matsuoka-Nakai constitutive
relation match our observations reasonably well for small angles,
$\vert \epsilon \vert < 0.1$, while the granular von Mises relation
predicts much too little circulation.  This suggests that the
Matsuoka-Nakai criterion may more accurately describe the yielding of
dense granular flows in the regime we are studying.

\begin{figure}
\centering
\includegraphics[width=3.375in]{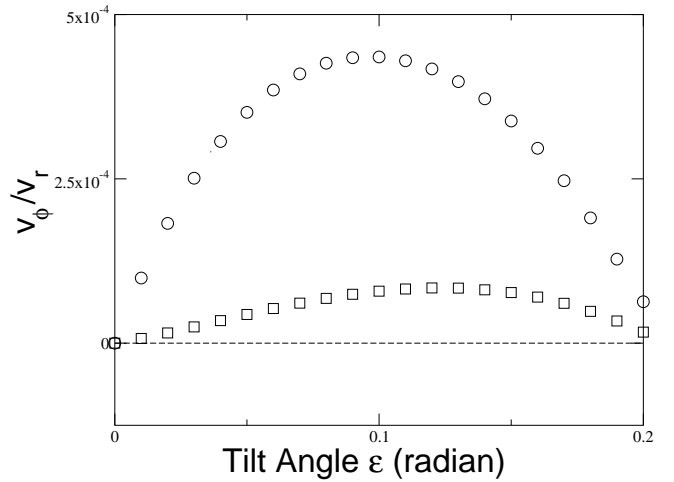}
\caption{When wall friction is extremely low, $\mu_w = 0.05$,
the numerically-predicted ratio of azimuthal to radial velocity as a
function of hopper tilt angle for both the Matsuoka-Nakai ($\circ$) and von Mises ($\square$) criteria is four orders of magnitude
smaller than the experimentally observed ratio at large tilt angles.}
\label{comparisonteflon}
\end{figure}

When the hopper wall friction is lowered (e.g. Figure~\ref{frictions}, teflon lining) the ratio $v_\phi/v_r$ is numerically predicted to be very small for both the Matsuoka-Nakai and von Mises yield criteria, as shown in Fig.~\ref{comparisonteflon}. Both sets of predictions are beneath the threshold of detection in this regime and we cannot distinguish between them experimentally.  We note, however, that we observe no trend for $\vert \epsilon \vert < 0.2$ in Fig.~\ref{frictions}.

\section{Velocity Fluctuations}

\begin{figure*} \centering \includegraphics[width=3.375in]{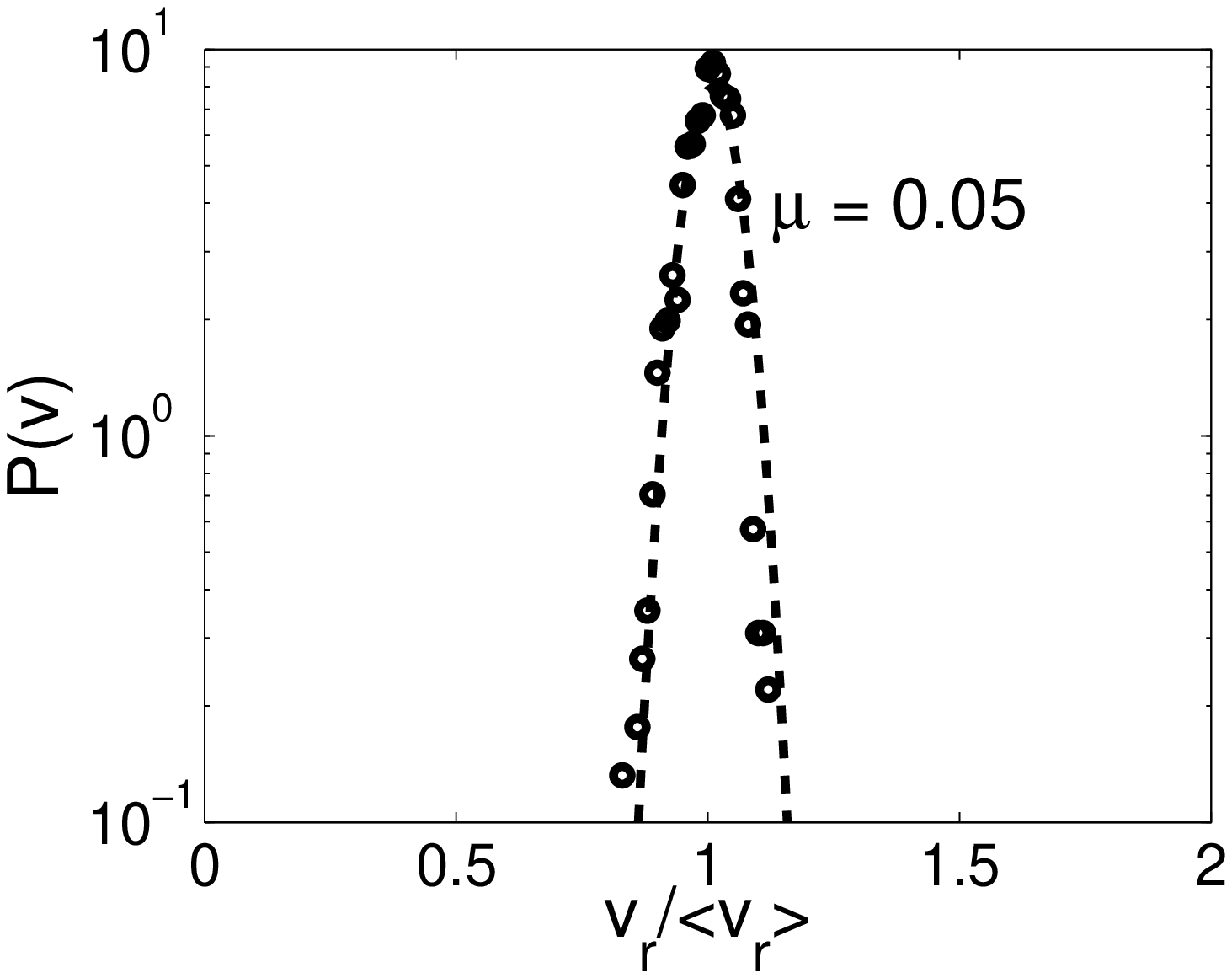}
\includegraphics[width=3.375in]{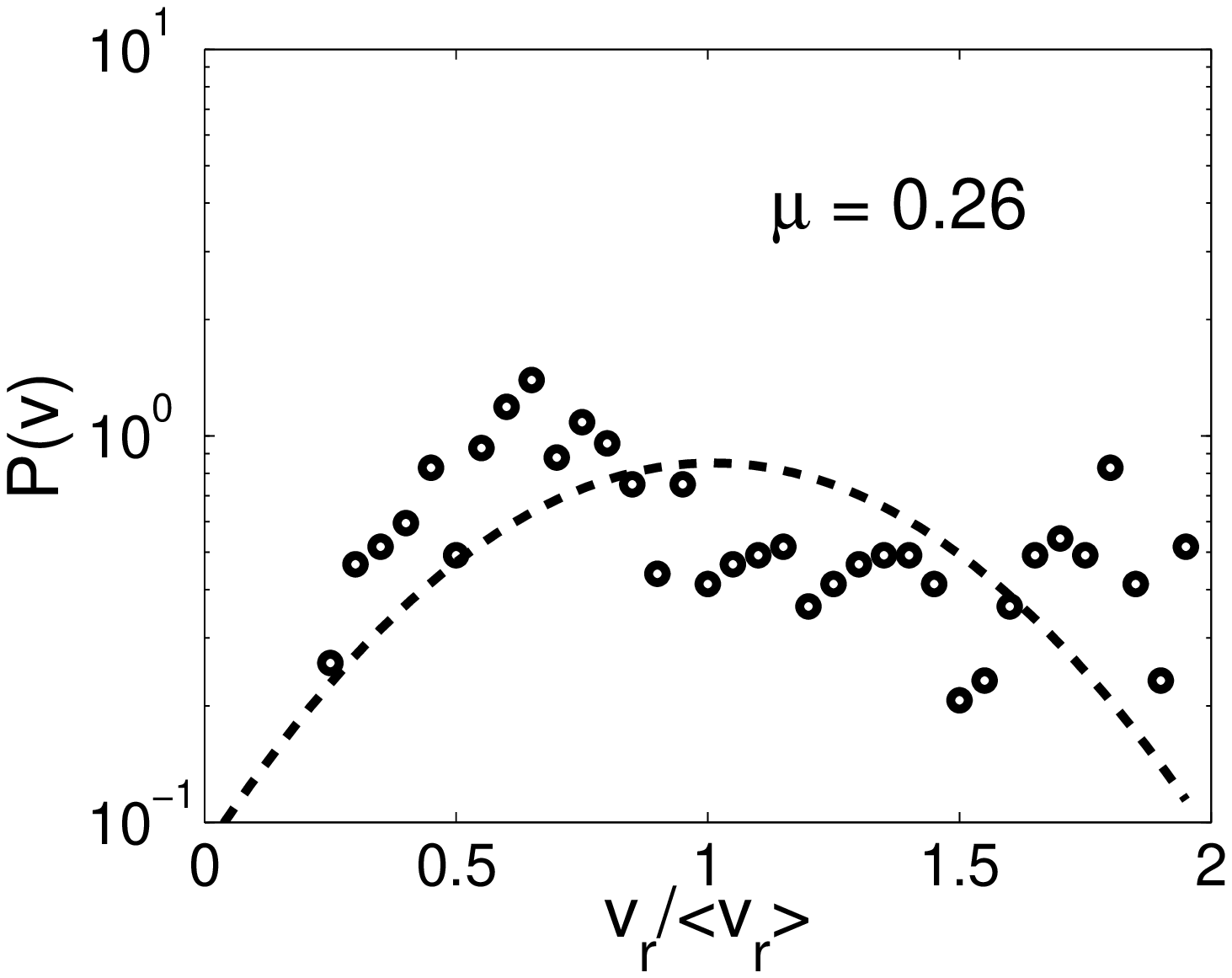}
\includegraphics[width=3.375in]{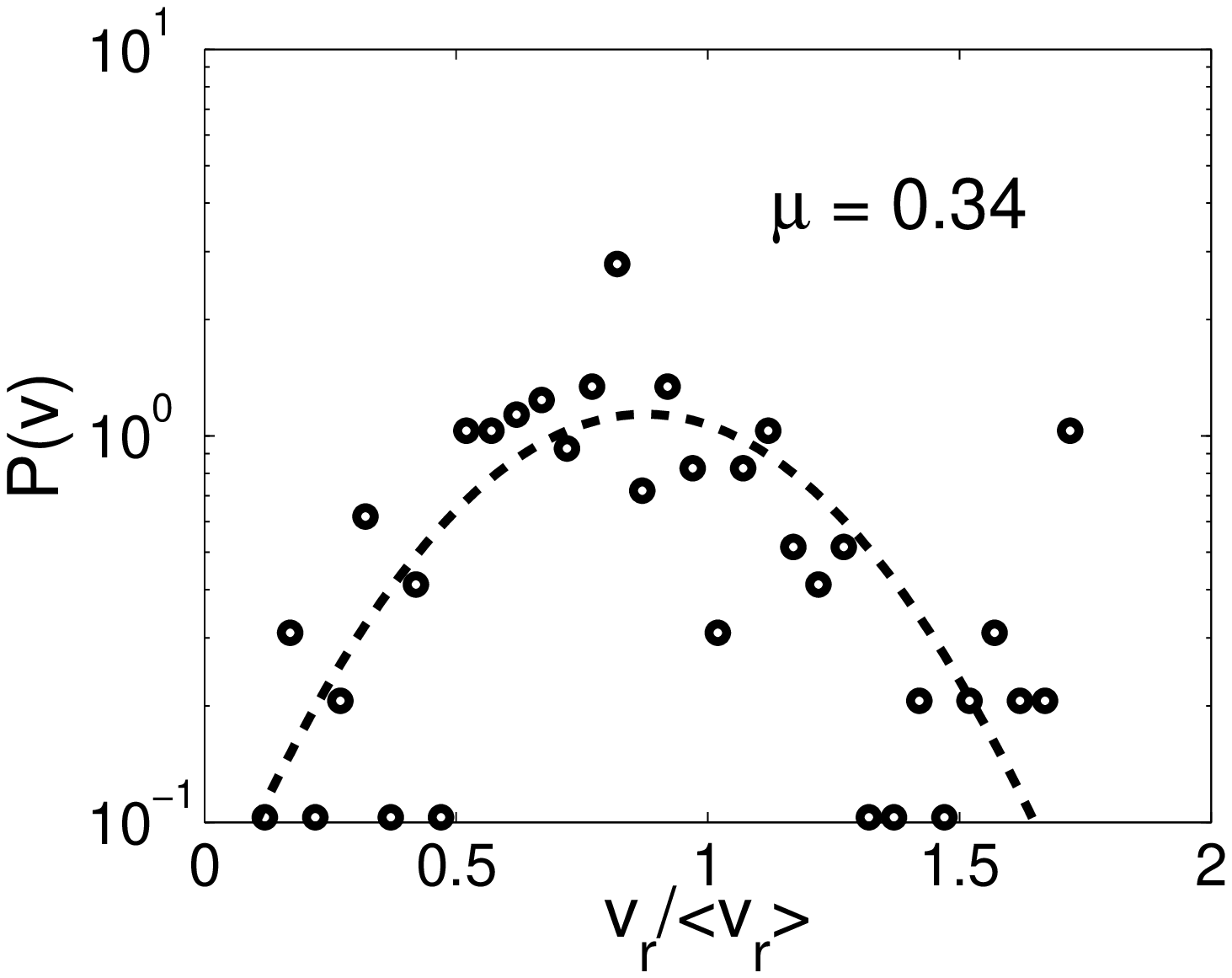}
\includegraphics[width=3.375in]{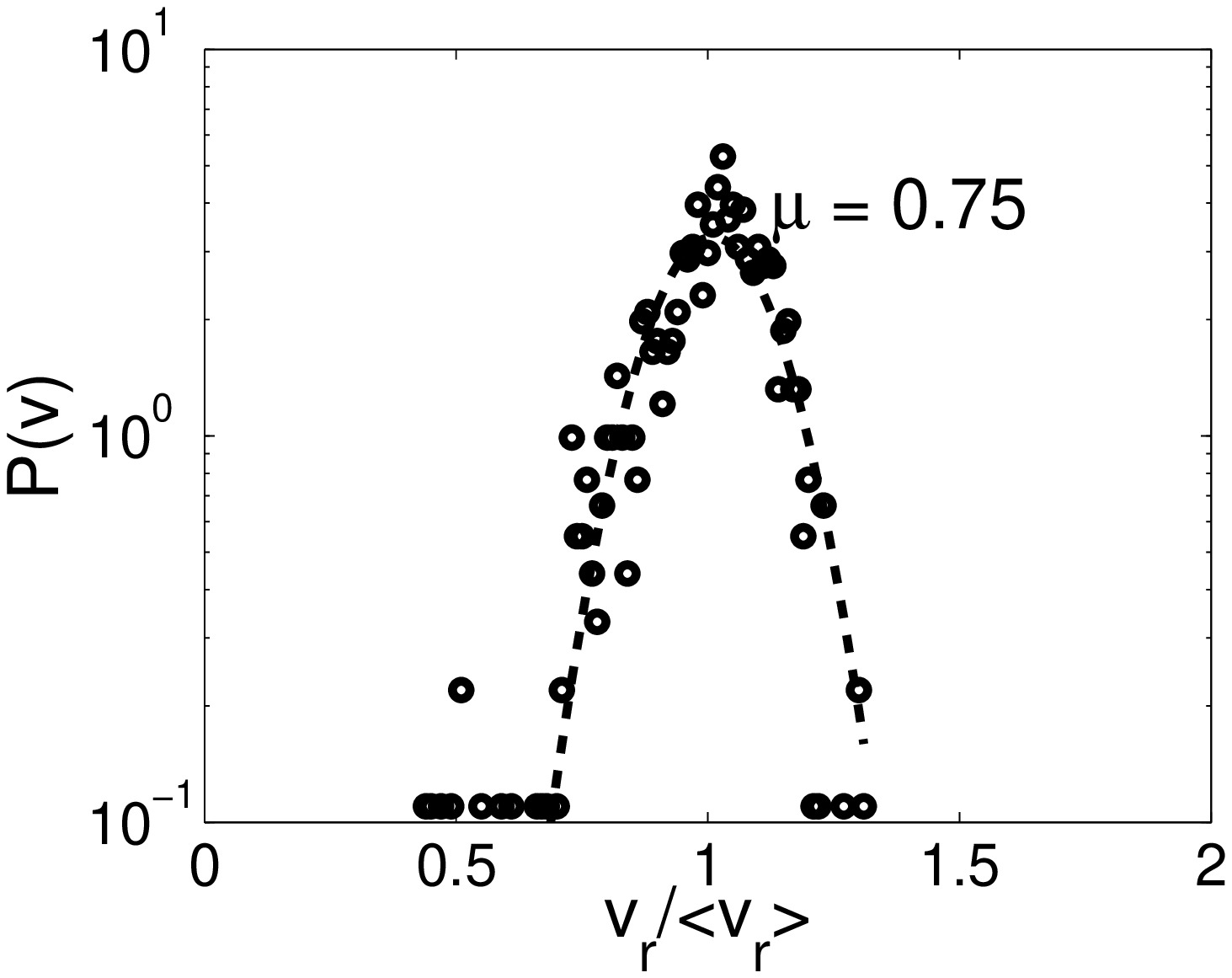} \caption{The width of the distributions of probability density for radial velocity varies non-monotonically with wall friction, with the extremes
having much narrower distributions than more moderate friction values. The distributions are
roughly fit by Gaussians ({\bf dashed lines}), with moderate values of friction having positive
skewness and extreme values having negative skewness.} \label{velocitydistributionsfric}
\end{figure*}

Although the numerical approach here describes the mean flow, we know
that stress in hoppers is subject to large fluctuations about the mean
\cite{Baxter93}.  Since Levy's flow rule assumes that the velocity in
our hopper is directly related to the stress, it is interesting to
examine any fluctuations in our velocity distributions
\cite{Jenike87}.

Previous work by Zhu and Yu using numerical Discrete Element Methods
has studied the probability distribution of velocities in a
flat-bottomed cylindrical hopper as wall friction was varied from $\mu
= 0.1$ to $0.5$ \cite{Zhu04}.  Although their cylindrical system had
notable differences from ours, including the presence of stagnant
regions and plug flow, separate distributions for the radial component
of flow were calculated, allowing rough comparison.  In the DEM
simulations it was found that the fluctuations decrease as wall
friction is reduced.

In Fig.~\ref{velocitydistributionsfric} we show the probability density 
distribution of radial velocity in an untilted hopper for the four
wall frictions we examined.  We generate our velocity histograms by
scaling several sets of data by the mean radial velocity for that set.
Since the mean velocity varied for some experiments, for each value of
wall friction we use the largest data set with consistent radial
velocities to generate our histograms.  Within one data set, the
velocity fluctuates sharply with time due to both our tracking
technique which can produce frames in which no velocities are assigned
and the large variability inherent to dense granular flows
\cite{Menon97,Menon06}.  The power spectrum for the time series does
not indicate periodicity.

As friction is increased from moderate values we observe a decrease in
the overall velocity fluctuations in agreement with the numerical
predictions.  The coefficient of friction for Teflon is below the
range studied previously, and in that case we observe that the
fluctuations are suppressed relative to moderate wall frictions,
indicating a non-monotonic dependence of velocity fluctuations on wall
friction.  All our observed distributions have skew, positive for
moderate friction and negative for extreme frictions, but because we
have studied velocities interpolated from many frames we do not see
the wide distributions of Moka and Nott \cite{Nott05}.

\begin{figure}
\centering
\includegraphics[width=3.375in]{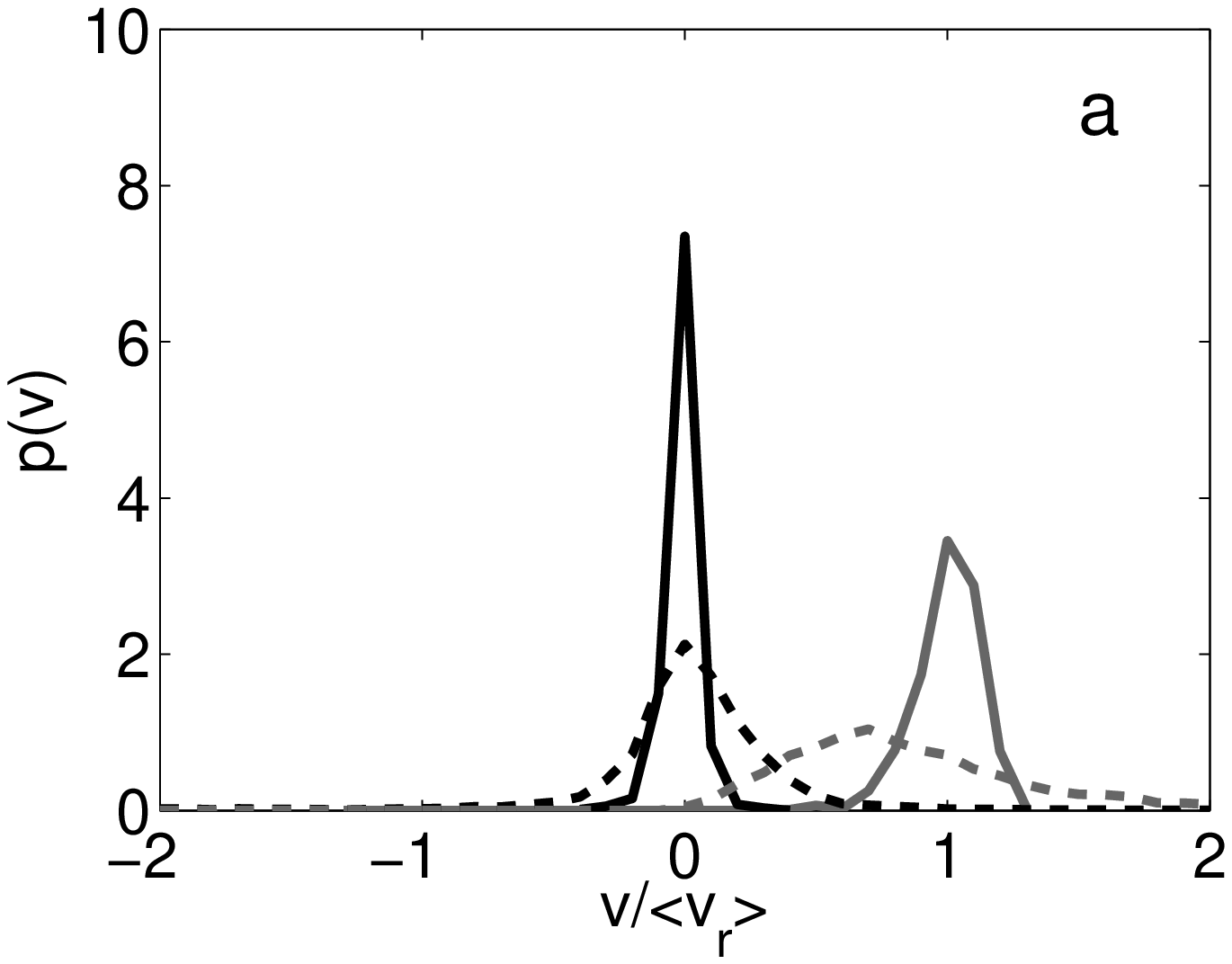}
\includegraphics[width=3.375in]{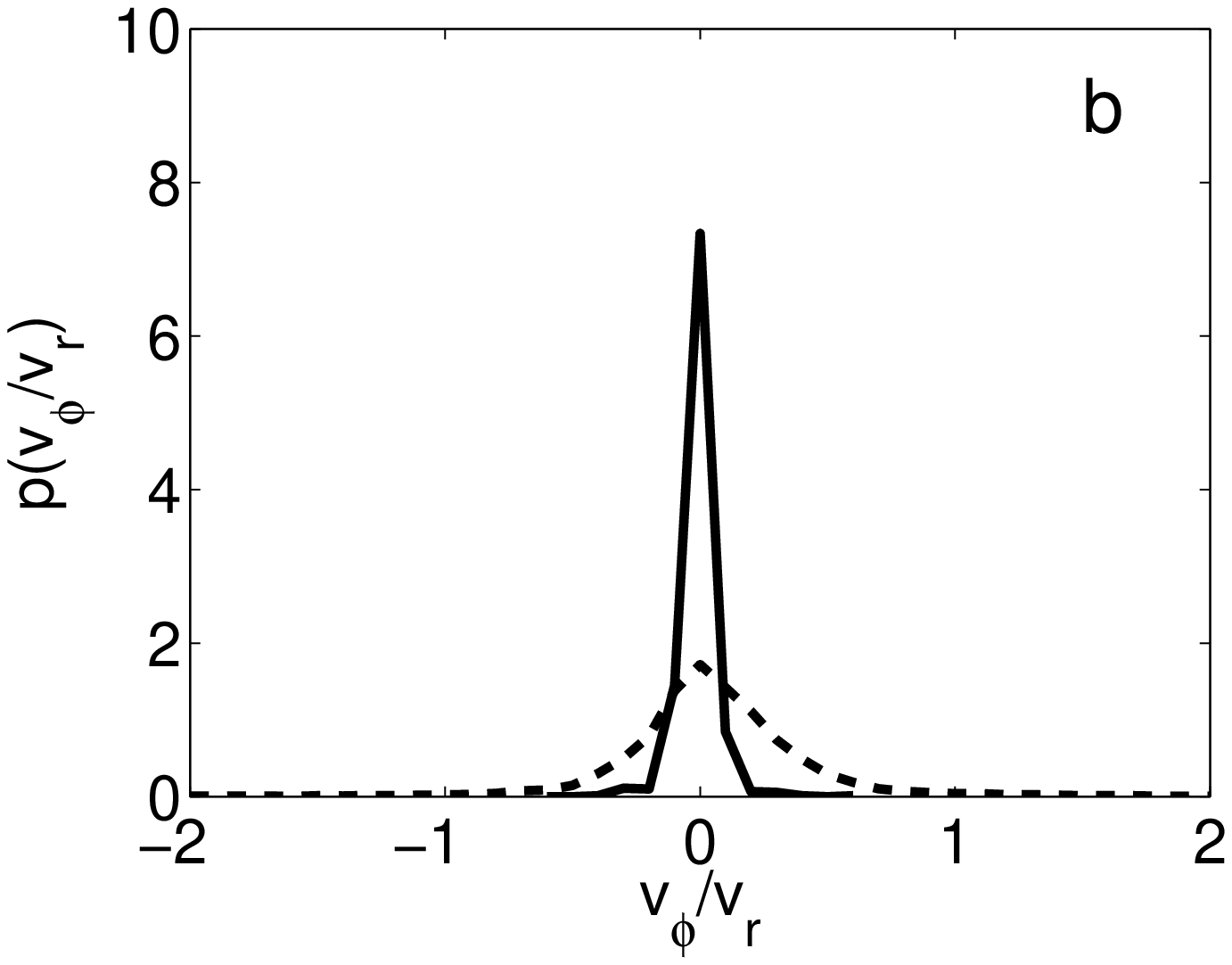}
\caption{Probability density distributions of {\bf a)} the azimuthal (left)
and radial (right) components of the velocity and {\bf b)} $\frac{v_\phi}{v_r}$ for
an untilted hopper ({\bf solid line}) and  a hopper at tilt $\epsilon = 0.14$ ({\bf dashed line}).  These results are for an
Emery paper-lined hopper, although they are representative of other
wall frictions.}
\label{velocitydistributionstilt}
\end{figure}

Finally, we examined the distribution of velocities as a function of
tilt angle.  For all wall linings, independent of the presence of
secondary circulation, we observed an increase in fluctuations around
the mean values as the tilt angle was increased.  Figure
\ref{velocitydistributionstilt} shows the probability distributions
for the radial and azimuthal components of the velocity as well as the
ratio of those two components for untilted and highly tilted cases.
For both tilted and untilted hoppers the means are similar, but the
distribution grows much wider with tilt angle.  Interestingly,
although the distributions are wider for the highly tilted hopper, the
distribution of $v_\phi/v_r$ is of roughly the same width as that of
$v_\phi$, indicating that the fluctuations of the two components are
correlated.  This is associated with the fact that individual grains
follow roughly fixed trajectories even if they are different from the
mean flow.  These persistent trajectories may indicate long-term
correlation of particle contacts that might be expected for clusters
of grains moving together, such as the ``granular eddies'' of
Erta\c{s} and Halsey \cite{Halsey02}.

\section{Conclusion}

We have observed that real three-dimensional flows of granular matter
are more complex than the idealized case of perfectly radial
Jenike-like flow.  In response to small perturbations (in the form of
tilts), non-radial velocities arise leading to potentially large
shifts in behavior.  For small tilts we have found that the magnitude
of circulation numerically predicted using the Matsuoka-Nakai constitutive
relation is closer to observations than numeric predictions made with the more
traditional, generalized von Mises condition.  At larger tilts we observe that the non-radial flow becomes even more pronounced, exceeding numerical predictions for all constitutive relations.

We have examined the influence of wall friction and discovered that
extreme values, low or high, act to suppress secondary circulation.
In the case of large perturbations and low friction, we see an abrupt
and unpredicted onset of secondary circulation in the opposite
direction of the moderate friction case.

Finally, we have observed that the fluctuations of individual grain
velocities grow as tilt is increased.  We find this to be true
independent of lining materials.  We also note that a single grain may
follow a trajectory that is quite different from the mean flow for
long periods of time, which produces large, correlated fluctuations.
Such particle-scale fluctuations are not captured by continuum models,
but do not seem to affect the mean behavior.

\section*{Acknowledgements}

JFW and RPB acknowledge funding from National Science Foundation
grants DMR-0137119, DMR0555431, and DMS-0204677 and NASA grant
NNC04GB08G.  PAG acknowledges funding from NSF Grants DMS-0244488 and
DMS-0410561.

\bibliography{../granular}
\end{document}